\documentclass[conference]{IEEEtran}

\ifCLASSINFOpdf
   \usepackage[pdftex]{graphicx}
\else
   \usepackage[dvips]{graphicx}
\fi

\usepackage{tikz}
\usepackage{multirow}
\usepackage{tabularx,booktabs}
\usepackage[font={footnotesize}]{caption}
\usepackage{pgfplots}
\usepackage{diagbox}
\usepackage{graphicx} 
\usepackage{epstopdf} \usepackage[font=small,labelfont=rm]{caption}
\usepackage[utf8x]{inputenc}
\usepackage{tabularx,ragged2e}
\usepackage[cmex10]{amsmath}
\usepackage{algorithm}
\usepackage{algpseudocode}
\usepackage{subcaption}
\usepackage{url}
\usepackage{siunitx}
\usepackage{pifont}
\usepackage[inline]{enumitem}
\DeclareSIUnit{\kWh}{kWh}
\usepackage[nobiblatex]{xurl}

\makeatletter

\let\old@ps@IEEEtitlepagestyle\ps@IEEEtitlepagestyle
\def\confheader#1{%
    \def\ps@IEEEtitlepagestyle{%
        \old@ps@IEEEtitlepagestyle%
        \def\@oddhead{\strut\hfill#1\hfill\strut}%
        \def\@evenhead{\strut\hfill#1\hfill\strut}%
    }%
    \ps@headings%
}

\makeatother

\confheader{%
    \parbox{20cm}{\textbf{\textit{Accepted to IEEE Belgrade PowerTech, 2023}}}
}

\begin{document}

\title{MARL-iDR: Multi-Agent Reinforcement Learning for Incentive-based Residential Demand Response}


\author{\IEEEauthorblockN{Jasper~van~Tilburg, Luciano~C.~Siebert,
~\IEEEmembership{Member,~IEEE,}
        and~Jochen~L.~Cremer
        ~\IEEEmembership{Member,~IEEE,}}
\IEEEauthorblockA{\emph{Faculty Electrical Engineering, Mathematics \& Computer Science} \\
\emph{Delft University of Technology} \\
Delft, The Netherlands
    \\\{L.CavalcanteSiebert, J.L.Cremer\}@tudelft.nl}
}

\maketitle

\begin{abstract}
This paper presents a decentralized Multi-Agent Reinforcement Learning (MARL) approach to an incentive-based Demand Response (DR) program, which aims to maintain the capacity limits of the electricity grid and prevent grid congestion by financially incentivizing residential consumers to reduce their energy consumption. The proposed approach addresses the key challenge of coordinating heterogeneous preferences and requirements from multiple participants while preserving their privacy and minimizing financial costs for the aggregator. The participant agents use a novel Disjunctively Constrained Knapsack Problem optimization to curtail or shift the requested household appliances based on the selected demand reduction. Through case studies with electricity data from $25$ households, the proposed approach effectively reduced energy consumption's Peak-to-Average ratio (PAR) by $14.48$\% compared to the original PAR while fully preserving participant privacy. This approach has the potential to significantly improve the efficiency and reliability of the electricity grid, making it an important contribution to the management of renewable energy resources and the growing electricity demand.
\end{abstract}





\section{Introduction}\label{sec:introduction}
Demand Response (DR) initiatives are promising to satisfy the increasing need for flexibility to prevent grid congestion due to growing demands and the intermittent nature of renewable energy resources \cite{Li2017DemandSystems}. 
DR programs can be either price-based, where the variation in the price policy influences the demand, or incentive-based, where companies offer electricity consumers financial incentives to reduce or shift their energy consumption.
Incentive-based DR (IBDR) programs are considered reward-wise programs, whereas price-based programs are considered punishment-wise programs. The voluntary nature of reward-wise programs makes people more positive and responsive in the long term. In contrast, the obligatory nature of the punishment-wise program makes people nervous, and responses are more transient \cite{Baboli2012CustomerModel}.
IBDR programs already contribute to flexible demands in the industrial sector. Still, much less in the residential sector \cite{Wang2020HowQuestionnaires}, which is a missed opportunity as residential consumers represent a significant share of electricity demand, e.g., almost half of the total energy consumption in the U.S. \cite{Annual2021}. Moreover, residential loads can provide a more reliable and continuous response than large industrial loads \cite{Asadinejad2016SensitivityElasticity}.
However, the participation of residential consumers in IBDR programs is challenging to realize since residential participants
\begin{enumerate*}[label=(\arabic*)]
    \item typically do not meet minimum levels of active load required to participate in the programs,
    \item may not be able to respond quickly to DR events and
    \item have higher privacy requirements than industrial consumers. 
\end{enumerate*}
First, existing IBDR programs are more suitable for residential participants when their loads are aggregated as a single participant in the IBDR program. Aggregators are key stakeholders in the electricity market, acting as an intermediary between the DSO and the consumer and creating the opportunity for residential consumers to participate in IBDR programs \cite{Zhou2018SmartManagement}.
Second, requests for load reductions in IBDR program may come up unannounced and require nearly real-time response, which residential participants may not be able to manage. One approach to solve this issue is to automate the response locally at the consumer via a Home Energy Management System (HEMS). 
Third, to preserve the privacy of residential participants, the aggregator does not have access to detailed information about the residents' preferences. 
Model-based approaches to automate IBDR programs are centralized and require exhaustive information about individual participants, which may not be available or may cause privacy issues \cite{Maharjan2013DependableApproach}\cite{Yu2017Incentive-basedApproach}. In addition, these centralized approaches rely on conventional optimization methods like linear programming \cite{Gholian2016OptimalGrid}\cite{Senevirathne2019OptimalEnvironment} or dynamic programming \cite{Zhang2018Profit-maximizingControl}, which make real-time computation infeasible for a large number of participants in the program. 

Reinforcement Learning (RL) is a promising approach for decision-making in IBDR programs since it does not require any information about the organization of the program or other participants (model-free)(see Appendix A). Second, it can control multiple agents, which allows for scaling up the number of participants. Third, once trained it can decide nearly instantly, facilitating future real-time control applications.   

This paper aims to answer the following research question: How can RL induce flexibility in residential demands to prevent grid congestion while preserving privacy and considering the heterogeneous preferences of residential consumers? This paper proposes a novel decentralized Multi-Agent Reinforcement Learning approach for Incentive-based DR (MARL-iDR) to answer the research question. The Markov Decision Process (MDP) is the guiding assumption to model sequential decision-making in IBDR programs. The proposed approach considers simultaneously a single Aggregator Agent (AA) and multiple participant agents aiming to maximize their rewards. The aggregator learns to deploy a suitable incentive based on one-step-ahead predictions of participant electricity demands, the target load reduction set by the DSO and participants' response to the incentive. The Participant Agent (PA) learns to respond to incentives by limiting consumption, which is achieved by shifting or curtailing household appliances, e.g. electric vehicles, dishwashers, and air conditioning, while preserving user satisfaction. The optimal power assignment is achieved through the proposed internal execution of a Disjunctively Constrained Knapsack Optimization (DCKP). This approach supports moving from inflexible centralized grid operation towards decentralized real-time automation while maintaining capacity limits and preserving consumer privacy and comfort with minimal information exchange.

The main contributions of this work are: 
\begin{itemize}
    \item An environment model that formulates an IBDR program, including an aggregator and multiple residential participants as an MDP. The environment model internally solves the DCKP to minimize participant dissatisfaction, taking the participant demand as input and schedules household appliances as output.
    \item MARL-iDR, a model-free MARL method for IBDR using deep Q-networks which makes real-time decisions for the aggregator and its residential participants, while preserving participants' privacy and accounting for heterogeneous preferences.
\end{itemize}

The rest of the paper is organized as follows. Section \ref{sec:related_work} discusses related work. The environment model is formulated as an MDP in Sec. \ref{sec:environment_model}. In Sec. \ref{sec:algorithm}, the MARL-iDR algorithm is described. In Sec. \ref{sec:case_study}, the results of a case study are presented to test the effectiveness of the approach. Finally, Section \ref{sec:conclusion} concludes the paper.

\section{Related Work}\label{sec:related_work}

Currently, much research is devoted to applying RL to DR. Some of those works focus on the industrial sector. \cite{Roesch2020SmartLearning} presents an approach to controlling a complex system of industrial production resources, battery storage, electricity self-supply, and short-term market trading using multi-agent RL.
\cite{Huang2019DemandApproach} present a deep RL-based industrial DR scheme for optimizing industrial energy management. To ensure practical application, they designed an MDP framework for industrial DR and used an actor-critic RL algorithm to determine the most efficient manufacturing schedule.

RL for DR in the residential sector has been proposed in numerous works. Many of these works focus on home energy management in a single household.
\cite{Wen2014OptimalLearning} presents an RL-based approach to DR for a single residential or small commercial building. They apply Q-learning with eligibility traces to reduce average energy costs by shifting the time of operation of energy consuming devices either by delaying their operation or by anticipating their future use and operating them at an optimal earlier time. The algorithm balances consumer dissatisfaction with energy costs and learns consumer choices and preferences without prior knowledge about the model.
\cite{Mathew2020IntelligentLearning} is a playful approach to residential DR using deep RL for scheduling loads in a single household. The authors propose an environment adapted from the Atari game Tetris where flexible blocks represent device loads. A DQN consisting of a convolutional network learns to schedule the load blocks.

Others focus on DR on the scale of the wholesale electricity market. \cite{Zhong2013CouponStudy} propose a voluntary incentive-based DR program targeting retail consumers with smart meters paying a flat electricity price. Load-serving entities provide consumers coupon incentives in anticipation of intermittent generation ramping and price spikes. Retail consumers' inherent flexibility is utilized while their base consumption is not exposed to wholesale real-time price fluctuations.
\cite{Yu2017Incentive-basedApproach}  propose an incentive-based DR model considering a hierarchical electricity market including grid operators, service providers or aggregators and small-load consumers. The proposed trading framework enables system-level dispatch of DR resources by leveraging incentives between interactors. A Stackelberg game is proposed to capture the interactions between interactors.

However, the previous approaches rely on model-based algorithms instead of model-free RL. The following works propose decentralized MARL methods for load scheduling of appliances in a collection of households.
\cite{Chung2020DistributedGrids} propose a model-free framework for scheduling the consumption profile of appliances in multiple households modelled as a non-cooperative stochastic game and apply RL to search for the Nash equilibrium. The authors emphasize the proposed method can preserve household privacy.
\cite{Zhang2019AScheduling} apply a cooperative RL approach to schedule controllable appliances of multiple households to minimise utility costs. The method performs explicit collaboration to satisfy global grid constraints. 
Both approaches emphasize the ability to scale with the number of participating households and to operate in real time.

These approaches, however, are price-based.
\cite{Lu2019Incentive-basedNetwork} proposes a real-time RL algorithm for incentive-based DR programs that supports service providers (aggregators) to purchase energy flexibility as a resource from its subscribed residential participants to balance energy fluctuations and enhance grid reliability. A single-agent RL is adopted to compute the close-to-optimal
incentive rates for heterogeneous participants. The participant's profit
and dissatisfaction are balanced with the service provider's objective.
\cite{Wen2020ModifiedModel} propose a similar method that includes PV generation and \cite{Xu2020ALearning} propose a similar method including historical incentives.

Research on applying RL for incentive-based residential DR is scarce, and the works that address this overlap either focus on home energy management or profits from the aggregator perspective. To address this research gap, this thesis proposes a multi-agent RL algorithm for DR that is incentive-based, residential and considers the interests of both the aggregator and multiple end consumers. A comparative overview is given in Table \ref{tab:related_work_overview}.

\begin{table}[]
    \centering
    \begin{tabular}{lcccc}
        \toprule
        \multirow{2}{*}{References} & \multirow{2}{*}{Residential} & {Incentive-} & {RL for the} & {RL for the} \\ 
        & & based & aggregator & consumer \\ \midrule
        \cite{Roesch2020SmartLearning}, \cite{Huang2019DemandApproach} &  & x &  &  \\ 
        \cite{Wen2014OptimalLearning}, \cite{Mathew2020IntelligentLearning} & x &  &  & x \\ 
        \cite{Zhong2013CouponStudy}, \cite{Yu2017Incentive-basedApproach} & x & x &  &  \\ 
        \cite{Chung2020DistributedGrids}, \cite{Zhang2019AScheduling} & x &  &  & x \\ 
        \cite{Lu2019Incentive-basedNetwork}, \cite{Wen2020ModifiedModel}, \cite{Xu2020ALearning} & x & x & x &  \\ 
        MARL-IDR & x & x & x & x \\ \bottomrule
    \end{tabular}
    \caption{Overview of related work and the aspects they consider. }
\label{tab:related_work_overview}
\end{table}

\section{Proposed Environment Model}\label{sec:environment_model}
\begin{figure}
    \centering
    \includegraphics[width=0.7\linewidth]{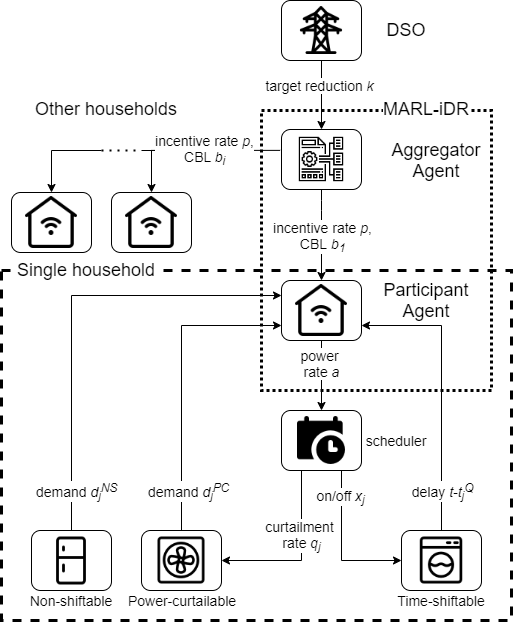}
    \caption{Architecture of the environment}
    \label{fig:architecture}
\end{figure}

The architecture for the proposed DR program is shown in Fig. \ref{fig:architecture}, and its components are described in detail throughout this section. The overall model considers multiple agents: one for the aggregator and one for each participant. The AA distributes incentives to the different PAs where each represents a residential household.  

\subsection{Assumptions for DR Program and Environment Model}
An assumption is that the DSO and the aggregator arrange a contractual agreement where the aggregator provides a continuous aggregated reduction in power consumption below a specified target in exchange for an agreed payment like in \cite{BaseBIP}. An additional assumption is that the participants only respond to the incentives offered in the IBDR program and not to fluctuations in the electricity price, as would be the case for price-based DR.

In IBDR programs, demand reductions are measured against a reference demand called the Customer Baseline Load (CBL).
The exact demand of participants in future time steps is unknown. Hence, aggregators have to estimate the demand of their participants.
Since residential participants show regular patterns in energy consumption throughout the day, the most prominent approach to demand estimation in the residential sector is based on historical consumption data.
In the proposed approach, the current day is matched to ten previous similar days, and the average consumption is taken considering changes in weather conditions as the CBL. Details for calculating the CBL are found in \cite{Sharifi2016CustomerEnvironment}.

The proposed environment model assumes an MDP, a framework for sequential decision-making, where a decision in a one-time step influences the next. The MDP is characterized by the Markov property, i.e. the state transitions depend solely on the current state of the environment and the current action taken. MDPs are described as a tuple $\langle S, A, P, R \rangle$ representing the state space, action space, transition probabilities and the reward function, respectively. One episode of an MDP consists of a finite sequence $T$ of discrete time steps $t$. The environment model assumes constant power consumption and incentive rates for a single time step. 
Based on these assumptions, RL agents observe the state of the environment in each time step, decide upon an action, and in return, receive a reward and transit to the next step and corresponding state. The agent considers immediate and future rewards multiplied by a discount factor $\gamma$. Therefore, the agent's objective is to maximize the cumulative discounted return \cite{Sutton2018ReinforcementEdition}. However, when the reward functions are equal for multiple agents, cooperation emerges \cite{Busoniu2010Multi-agentOverview}. This feature is interesting to explore for multiple agents in IBDR programs.

\subsection{Participant Agent}\label{sec:participant_model}
The set of all households in the DR program is $\mathcal{H}$. Each PA $i \in \mathcal{H}$ can control a set of appliances $\mathcal{D}_i$.
In practice, the PA could be integrated into a HEMS connected to smart meters and smart plugs to access the consumption measurements of household appliances. 
The objective of the PA is to approach the optimal balance between maximizing financial earnings and minimizing user dissatisfaction caused by curtailing or delaying appliances. The environment model for each household is defined by its state, the actions, rewards and the scheduler.

\subsubsection*{State}
The appliances $\mathcal{D}_i$ are divided into three subsets: time-shiftable appliances $\mathit{TS}$, power-curtailable appliances $\mathit{PC}$ and non-shiftable appliances $\mathit{NS}$ such that $\mathcal{D}_i = \mathit{TS}_i \cup \mathit{PC}_i \cup \mathit{NS}_i$.
The residents may submit an initial request to turn on appliance $j \in \mathcal{D}_i$ at time step $t^{I}_{i,j}$.

\begin{itemize}
    \item Time-shiftable appliance $\mathit{TS}$ are either on with constant power consumption or off. Time-shiftable appliances can be interruptible (e.g. EVs, where the charging can continue later) or non-interruptible, (e.g. washing machines and dishwashers that need to complete their washing programs without interruption). Their state is determined by the difference between the current time step $t$ and the initial request time steps $t^{I}_{j}$, i.e. the current delay $t - t^{I}_{j}$. 
    \item Power-curtailable appliances $\mathit{PC}$ allow to lower power consumption but do not allow a delay of usage (e.g. changing the setpoint of an AC, dimming lighting systems). Their state is the variable power demand in kilowatts $d^{\mathit{PC}}_j$. 
    \item Non-shiftable appliances $\mathit{NS}_i$ must run at all times without delay or curtailment. These appliances share a single state defined as the total power demand in kilowatts $d^{\mathit{NS}}_j$
\end{itemize}

The state $s_{t,i}$ of a household $i$ at the time step $t$ combines the information of all appliances $j \in \mathcal{D}_i$. In addition, the PA observes the incentive rate $p_t$ passed from the AA and its own projected CBL $b_{t,i}$. 
Hence, the observation of PA $i$ in time step $t$ is $o^{PA}_{t,i} = \{s_{t,i}, b_{t,i}, p_t\}$. \footnote{For the remainder of this subsection, the subscripts $t$ and $i$ are dropped as all equations apply to time step $t$ and household $i$.}

\subsubsection*{Action}
This paper proposes an action space of discrete power rates combined with an appliance scheduling optimization to ensure scalability in the number of appliances.
The problem is that when appliances are controlled directly by RL the action space for time-shiftable appliances increases exponentially with the number of appliances, i.e. the binary combination of time-shiftable appliances (either on or off) is $\mathcal{O}(2^{|\mathit{TS}_i|})$.
This problem of scalability is even more pressing for power-curtailable appliances where discretization in a set of $m$ levels of power consumption results in a combination growing with $\mathcal{O}(m^{|\mathit{PC}_i|})$.

To address this issue, a fixed action space is proposed that consists of discrete power rates $a \in A^{PA}$, which is a fraction of the total demand. Subsequently, the scheduler described in Section \ref{sec:scheduler} matches the appliances to the limit $l = a \cdot d$ where $d$ is the total appliances' demand. The resulting total power consumption may be lower than the limit $e \leq l$.

\subsubsection*{Reward}
The reward for the PA consists of two components \begin{enumerate*}[label=(\arabic*)]
    \item the financial reward for receiving incentives
    \item the dissatisfaction cost for preserving the satisfaction of the residents.
\end{enumerate*}
First, as the AA offers the PA incentive rate $p$ to reduce demand, the PA receives a financial reward when the total consumption $e$ is smaller than the CBL $b$. The financial reward $u$ paid from AA to PA is
\begin{equation}\label{eq:participant_incentive}
    u = p \cdot \max \, (0, \, b - e),
\end{equation}
\noindent
As the DR program is incentive-based (reward-wise) not price-based (punishment-wise), participants are not punished for consuming more than CBL $b$, i.e. they can only earn money, not lose anything.

Second, curtailing or shifting requested appliances causes dissatisfaction to the residents. In the case of time-shiftable appliance $j \in \mathit{TS}$, dissatisfaction cost $c_{j}^{\mathit{TS}}$ is a convex function of the delay
\begin{equation}  \label{eq:CTS}
    c_{j}^{\mathit{TS}} = \beta_{j} \Big(t + 1 - t^{I}_{j}\Big)^2 \quad \forall j \in \mathit{TS},
\end{equation}
\noindent 
Shifting appliance $j$ to time step $t+1$ instead of turning it on in time step $t$ means a delay of $t+1 - t^{I}_{j}$.
This function assumes the residents get increasingly dissatisfied when waiting longer for the appliance to run \cite{Xu2020AManagement}. In the case of power-curtailable appliance $j \in \mathit{PC}$, the dissatisfaction cost is a convex function of the power curtailment.
\begin{equation} \label{eq:CPC}
    c^{\mathit{PC}}_{j} = \beta_{j} \Big(\frac{1}{m} \cdot q_{j} \cdot d_{j}\Big)^2 \quad \forall j \in \mathit{PC},
\end{equation}
\noindent
where $q_j \in \{0, 1, \dots, m\}$ is a categorical variable corresponding to the power curtailment level. This function assumes residents get increasingly dissatisfied with increased curtailment \cite{Fahrioglu2001UsingModels}. $\beta_{j}$ is an appliance-specific dissatisfaction coefficient describing the tolerance of the residents for delay or power curtailment. In practice, this coefficient is a parameter that the residents can update in the HEMS according to their preferences. 

The total reward function combines financial reward $u$ and dissatisfaction cost $c$ as follows
\begin{equation}
    r^{PA} = u - \sum_{j \in \mathcal{D}} c_j
\end{equation}

\subsubsection*{Scheduler}\label{sec:scheduler}
As part of the PA for household $i$, the scheduler determines the optimal assignment of power to the appliances based on the overall demand limit $l$. The scheduler is a combinatorial optimization formulated as DCKP \cite{BenSalem2016OptimizationProblem}:
\begin{align}
    \text {minimize} \quad & \sum_{j \in \mathit{PC}} c_j^{\mathit{PC}} + \sum_{j \in \mathit{TS}} (1-x_j) \cdot c_j^{\mathit{TS}} \\
    \text{subject to} \quad & \sum_{j \in \mathit{PC}} \frac{1}{m} \cdot q_j \cdot d_j +  \sum_{j \in \mathit{TS}} x_j \cdot d_j \leq l \\
    & x \in \{0, 1\}, \, q \in \{0, 1, \dots, m\}
\end{align}
\noindent
where $x_j$ is a binary variable for each time-shiftable appliance $j \in \mathit{TS}$ corresponding to switching the appliance off ($x_j = 0$) or on ($x_j =1$). The optimization minimizes the total dissatisfaction from all appliances. The dissatisfaction costs from time-shiftable appliances $c_j^{\mathit{TS}}$ and power-curtailable appliances $c_j^{\mathit{PC}}$ are parameters computed with Eq. \eqref{eq:CTS} and Eq. \eqref{eq:CPC}. 
After solving the DCKP, the overall power demand is
\begin{equation}
    e = \sum_{j \in \mathit{PC}} \frac{1}{m} \cdot q_j \cdot d_j +  \sum_{j \in \mathit{TS}} x_j \cdot d_j
\end{equation}

\subsection{Aggregator Agent}\label{sec:aggregator}

\subsubsection*{State}
The state space of the AA is $o^{AA}_t = \{d_t, k\}$, where $d_t$ is the aggregated demand of all households $i \in \mathcal{H}$ and $k$ is the target reduction set by the DSO.

\subsubsection*{Action}
In each time step the AA selects an incentive rate $p_t$ to realize power reduction by the PAs.
The AA selects $p_t$ out of an action space of discrete incentives $A^{AA}$ in cents per kilowatt of demand reduction. 

\subsubsection*{Reward}
The reward of the AA is
\begin{equation}\label{eq:rAA}
    r^{AA}_t = -\bigg(\rho \cdot e^{+}_t + (1-\rho) \cdot \sum_{i \in \mathcal{H}} u_{t,i}\bigg),
\end{equation}
\noindent 
where the first term defines a penalty for exceeding target $k$ as $e^{+}_t = max(0, e_t - k)$. The second term is the total incentive paid to the PAs as defined in Eq. \eqref{eq:participant_incentive}. The trade-off between the two terms is determined by weighting factor $\rho$. Note that the AA is not rewarded for aggregated consumption below the target as it aims to reduce consumption to contribute to the DSO's capacity constraints, but not further reduce energy consumption.

\section{Proposed MARL-iDR Algorithm}\label{sec:algorithm}

The proposed MARL-iDR is a multi-agent algorithm where the AA and PA have indirectly opposing reward functions, i.e. actions in favour of the AA may have a negative influence on the reward of the PA and vice versa.
All MARL-iDR agents are trained simultaneously, hence, the agents deal with a moving target where the optimal policy changes as opposing agents change their policies. Simultaneous learning leads to non-stationary problems which invalidate most of the single-agent RL theoretical guarantees, e.g. the guarantee of convergence \cite{Busoniu2010Multi-agentOverview}. Despite these limitations, simultaneous learning has found numerous applications because of its simplicity \cite{Crites1995ImprovingLearning}\cite{Mataric1995LearningSystems}.

MARL-iDR effectively trades off exploration with exploitation (a fundamental concept in RL). MARL-iDR uses the action-selection strategy of $\epsilon$-greedy with decay, i.e. with probability $\epsilon$ the agent selects a random action where $\epsilon$ decreases over time with decay rate $\delta$ \cite{Caelen2007ImprovingAlgorithms}. With this strategy, MARL-iDR benefits from extensive exploration early in training and refinement of the policy in later stages.

MARL-iDR uses Deep Q-Networks (DQNs) to account for huge and continuous state spaces that are infeasible to Q-learning \cite{Sutton2018ReinforcementEdition}. DQN is a state-of-the-art Deep RL approach that estimates the Q-value of state-action pairs by means of a neural network $\theta$. For more stable training with DQNs, two features are used:
\begin{enumerate*}
    \item A separate target network $\theta^T$ for setting the target values to avoid non-stationary targets, while the original network is used for predicting Q-values.
    \item Experience replay to avoid the agent from forgetting previous experiences.
\end{enumerate*}
All experiences are saved in a replay buffer $B$. Instead of training the network only on the most recent experience, the network is trained on randomly sampled batches of experience from $B$.

The training procedure for MARL-iDR is Algorithm \ref{alg:training}.
At the start of the procedure a policy network and target network are initialized for each individual agent. Then, all agents train the networks for a number of episodes where each episode corresponds to a single day which has a sequence of $T$ time steps. In each time step $t$, first the AA selects an action. Next, each individual PA $i$ selects an action and immediately receive their reward. Finally, after all PAs decided their response to the AA the reward for the AA can be calculated. The training procedure takes a significant amount of time to learn policies for each agent, however, once trained, the RL agent can be deployed in real-time using policy $\pi$:
\begin{equation}
    \pi(s) = \underset{a}{argmax} \, Q(s, a \, | \, \theta)
\end{equation}

\begin{algorithm}
    \caption{MARL-iDR training procedure}\label{alg:training}
    \begin{algorithmic}
        \State Initialize $\theta^{AA}, \theta^{T,AA}, B^{AA}$
        \State Initialize $\theta_{i}^{PA}, \theta_{i}^{T,PA}, B_{i}^{PA} \quad \forall i \in \mathit{H}$
        \State Initialize $\epsilon_0 \gets 1.0$
        \State Initialize $\delta, \quad 0 <  \delta < 1$

        \ForAll{episodes}
            \State Initialize target reduction $k$
            \State Initialize rewards $r^{AA}_{0}, r^{PA}_{0} \gets 0$
            \State $\epsilon_{t} = \epsilon_{t-1} \cdot \delta$
            \ForAll{time steps $t$}
                \State Predict demand $d_t$ and set CBLs $b_t$
                \State $o^{AA}_t \gets \langle d_t, k \rangle$
                \State Add $\langle o^{AA}_{t-1}, p_{t-1}, r^{AA}_{t-1}, o^{AA}_t \rangle$ to $B$
                \State Train $\theta^{AA}$ given $B$
                \State Select $p_t$ using $\epsilon$-greedy
                \ForAll{PAs $i \in \mathit{H}$}
                    \State Observe state of appliances $s_{t,i}$ and compute $c_{t,i}$
                    \State Observe CBL $b_{t,i}$, incentive rate $p_t$
                    \State $o^{PA}_{t,i} \gets \langle s_{t,i}, b_{t,i}, p_t \rangle$
                    \State Add $\langle o^{PA}_{t-1, i}, a_{t-1, i}, r^{PA}_{t-1, i}, o^{PA}_t \rangle$ to $B$
                    \State Train $\theta_{i}^{PA}$ given $B$
                    \State Select $a_{t,i}$ using $\epsilon$-greedy
                    \State Obtain $x_{t,i}$ and $q_{t,i}$ by solving DCKP \\ \hspace{2cm} with input: $d_{t,i}, c_{t,i}, a_{t,i}$
                    \State Update $s_{t+1,i}$ according to $x_{t,i}$ and $q_{t,i}$
                    \State Calculate PA reward $r^{PA}_{t, i}$
                \EndFor
                \State Calculate AA reward $r^{AA}_t$
            \EndFor
        \EndFor
    \end{algorithmic}
\end{algorithm}

\section{Case study}\label{sec:case_study}
The case study tests the effectiveness of the environment model and the MARL-iDR algorithm taking four aspects into consideration:
\begin{enumerate*}
    \item The policies learned by the agents
    \item Information exchange and if the privacy of the participants is preserved
    \item Computational efficiency and finally
    \item Economics for the aggregator considering a varying weighting factor $\rho$.
\end{enumerate*}

\subsection{Simulation data and test setup} \label{sec:sim_data}

This case study uses appliance requests and consumption data from 25 real-world households from the PecanStreet dataset \cite{PECANInc.}. The dissatisfaction coefficients $\beta_{j}$ for the appliances are sampled from a normal distribution to introduce heterogeneity to the households. The type, demand and dissatisfaction coefficients of the appliances selected for the simulations are in Table \ref{tab:appliance_specs}.

\begin{table}[]
\centering
\begin{tabular}{lcccc}
\toprule
\multirow{2}{*}{Appliance}  & \multirow{2}{*}{Type} & Demand  & \multicolumn{2}{c}{Dissatisfaction coeff.} \\
&  &(kW) & mean & std \\ \midrule
Dryer & \begin{tabular}[c]{@{}l@{}}$\mathit{TS}$, $\mathit{NI}$ \end{tabular} & $2.0$ & $0.2$ & $0.2$ \\ 
Washing  machine (WM)                                                                           & $\mathit{TS}$, $\mathit{NI}$  & 1.0 & 0.1 & 0.1 \\ 
Dishwasher (DW) & $\mathit{TS}$, $\mathit{NI}$  & 2.0 & 0.06 & 0.05 \\ 
EV  & $\mathit{TS}$, $\mathit{I}$ & 4.0 & 0.04 & 0.05 \\ 
AC & $\mathit{PC}$   & 0 - 4.0 & 3.0 & 1.0           \\ 
Non-shiftable & $\mathit{NS}$ & 0 - 5.0 & - & - \\ \bottomrule
\end{tabular}
\caption{Household appliances and their parameters, e.g, non-interruptible ($\mathit{NI}$) and interruptible ($\mathit{I}$). 
}
\label{tab:appliance_specs}
\end{table}

The  MARL-iDR algorithm is trained for $5000$ episodes, discretized in $T=96$ time steps of $\SI{15}{min}$ and randomly sampled from the training period April 1, 2018, to October 31, 2018. The action space for the PAs and the AAs are defined as $A^{PA} = \{0.0, 0.1, ..., 1.0\}$ and $A^{AA} = \{0, 1, ..., 10\}$ respectively. The scheduler selected from $m=10$ curtailment levels.
The AA and PAs have an individual DQN with a learning rate $\eta=0.001$. The discount rate $\gamma=0.9$ and $\epsilon$-decay rate $\delta=0.999$. Weighting factor $\rho$ is $0.5$. Finally, the target reduction $k$ is defined at $80\%$ of the peak demand. The algorithm is available online in a GitHub repository \cite{van_Tilburg_Case_study_for}. The algorithm is validated on each day in July. The simulations were conducted on a 2.20 GHz, Intel 6-core i7-8750 CPU with $16$ GB RAM, running Windows $10$.

A baseline was used to compare the performance of the proposed MARL-iDR. The baseline considered the optimal myopic action per time step (i.e. not considering future rewards). In other words, PAs selected the best action such that $a^*_i = argmax \, \{r^{PA}_i|p\}$, and the AA selected the optimal incentive defined by $p^* = argmax \, \{r^{AA}|a^*_i, \, \forall i \in H\}$. This myopic baseline requires full model knowledge and can only consider immediate rewards (short-sighted). 

\subsection{Load reductions and incentive rates}
\begin{table}
\centering
\begin{tabular}{lccc}
\toprule
& \multirow{2}{*}{No DR} & \multirow{2}{*}{MARL-IDR} & {Myopic}\\ 
& & & baseline \\
\midrule
{Peak load (kW)}          & 86.25  & 74.39  & 69.23           \\ 
{\begin{tabular}[c]{@{}l@{}}Mean load (kW)\end{tabular}}          & 47.80  & 45.37  & 46.23           \\ 
{PAR}                                                                & 1.80   & 1.64   & 1.50           \\ 
{Surplus consumption (kWh)} & 35.79  & 3.93   & 0.49            \\ 
{Total incentive (¢)} & 0.0  & 2122   & 1917            \\ 
{Average  dissatisfaction cost} & 0.0  & 17.36   &  12.26           \\ 
{Average incentive income (¢)} & 0.0  & 84.88   &  76.68           \\ \bottomrule\end{tabular}
\caption{Results averaged per day in July.}
\label{tab:results}
\end{table}

\begin{figure}
\begin{subfigure}{0.4\textwidth}
\begin{tikzpicture}
\begin{axis}[
        axis on top,
        width=\textwidth,
        height = 6em,
        scale only axis,
        enlargelimits=false, 
        ytick={0,2,4,6},     
        xtick={12,14,16,18,20,22,24}, 
        ylabel={Incentive (\textcent)},
        xlabel={Time(h)},
        xmin=12,
        xmax=24,
        ymin=0,
        ymax=6,
        y label style={at={(axis description cs:0.1,.5)},anchor=south}, 
        ]        
	\addplot graphics[xmin=12,ymin=0,xmax=23.75,ymax=6] {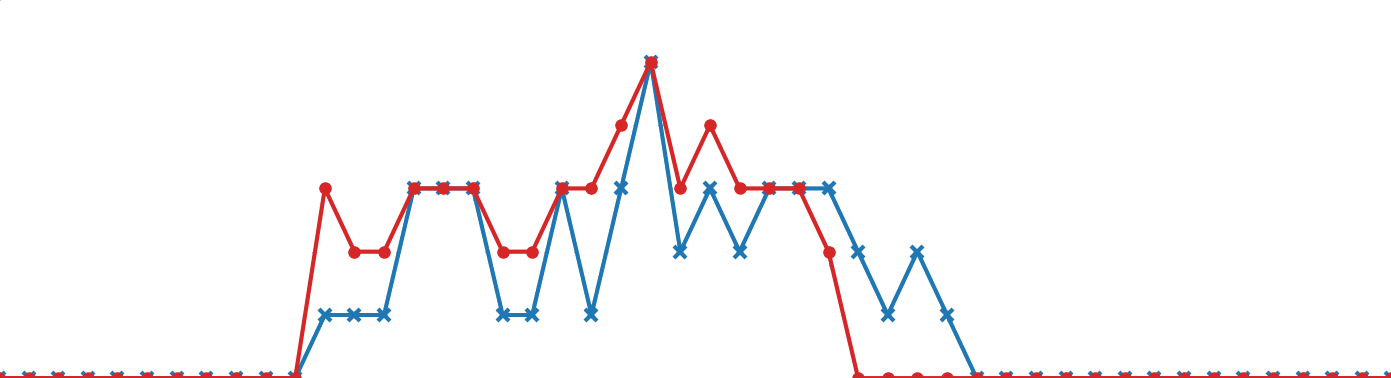};
  \end{axis}
\end{tikzpicture}
   \caption{Incentive rates selected by the AA.}
\label{fig:results_incentive}
\end{subfigure} 
\begin{subfigure}{0.4\textwidth} 
\begin{tikzpicture}
\begin{axis}[
        axis on top,
        width=\textwidth,
        height = 6em,
        scale only axis,
        enlargelimits=false, 
        ytick={40,60,80,100},     
        xtick={12,14,16,18,20,22,24}, 
        ylabel={Power (kW)},
        xlabel={Time(h)},
        xmin=12,
        xmax=24,
        ymin=40,
        ymax=105,
        y label style={at={(axis description cs:0.1,.5)},anchor=south}, 
        ]        
	\addplot graphics[xmin=12,ymin=40,xmax=23.75,ymax=105] {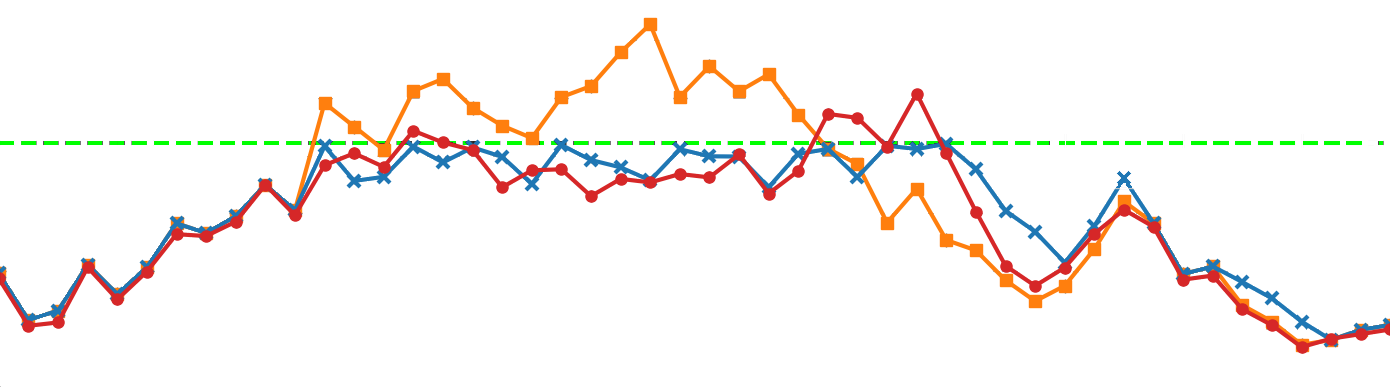};
  \end{axis}
\end{tikzpicture}
   \caption{Aggregated load curve with target reduction (\protect\includegraphics[height=0.5em]{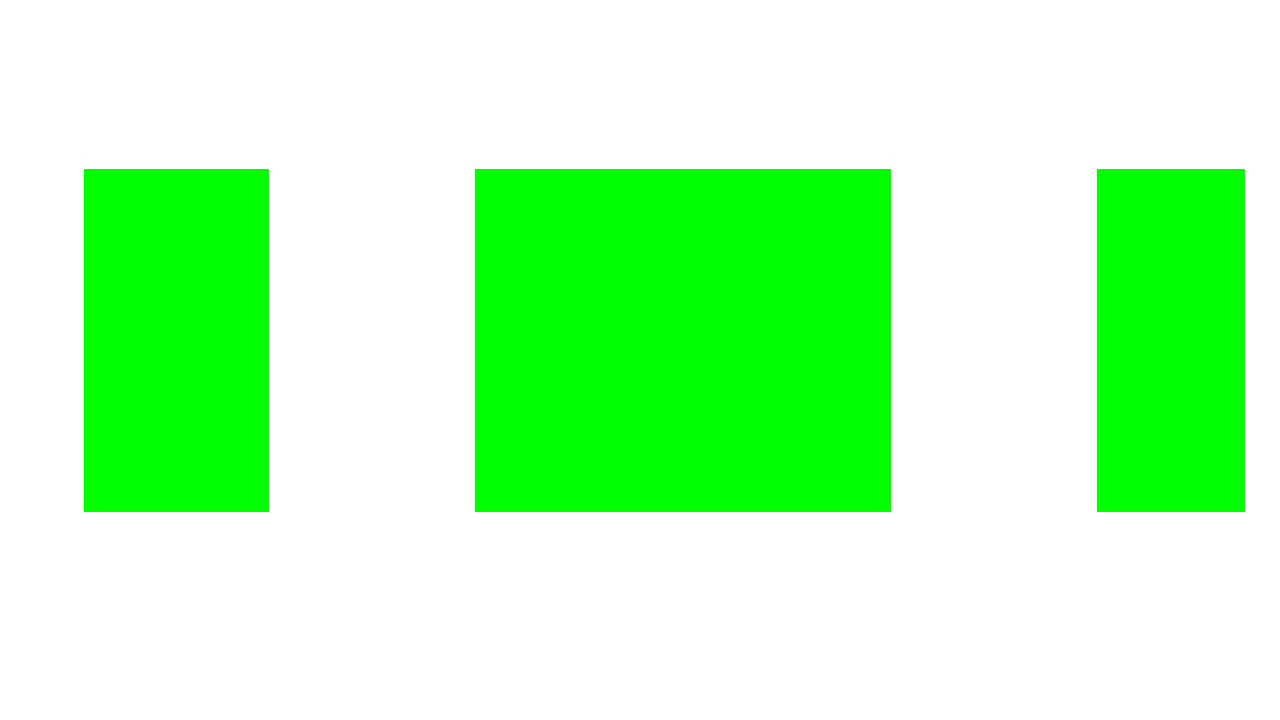}).}
\label{fig:results_aggregated_load}
\end{subfigure}
\begin{subfigure}{0.4\textwidth} 
\begin{tikzpicture}
\begin{axis}[
        axis on top,
        width=\textwidth,
        height = 6em,
        scale only axis,
        enlargelimits=false, 
        ytick={0,2.5,5,7.5,10},     
        xtick={12,14,16,18,20,22,24}, 
        ylabel={Power (kW)},
        xlabel={Time(h)},
        xmin=12,
        xmax=24,
        ymin=0,
        ymax=10,
        y label style={at={(axis description cs:0.1,.5)},anchor=south}, 
        ]        
	\addplot graphics[xmin=12,ymin=0,xmax=23.75,ymax=10] {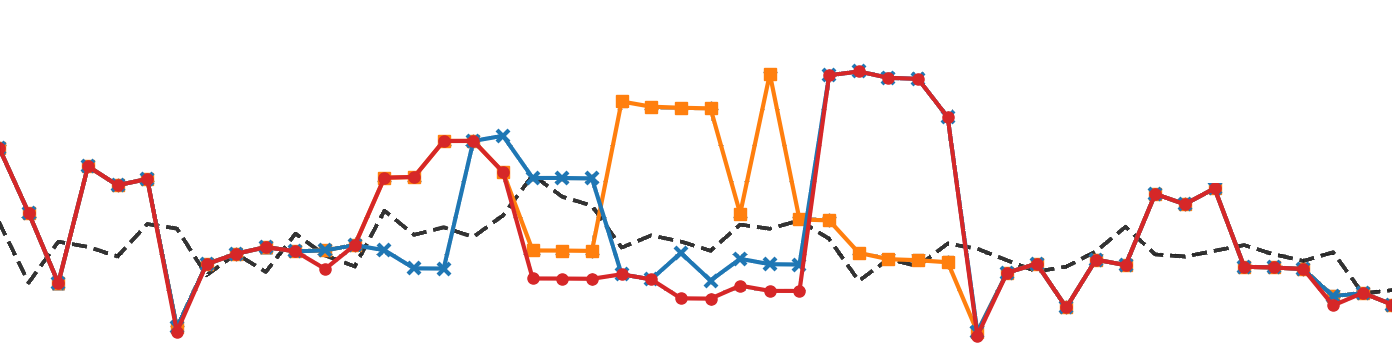};
  \end{axis}
\end{tikzpicture}
   \caption{Load curve of an individual household with CBL (\protect\includegraphics[height=0.4em]{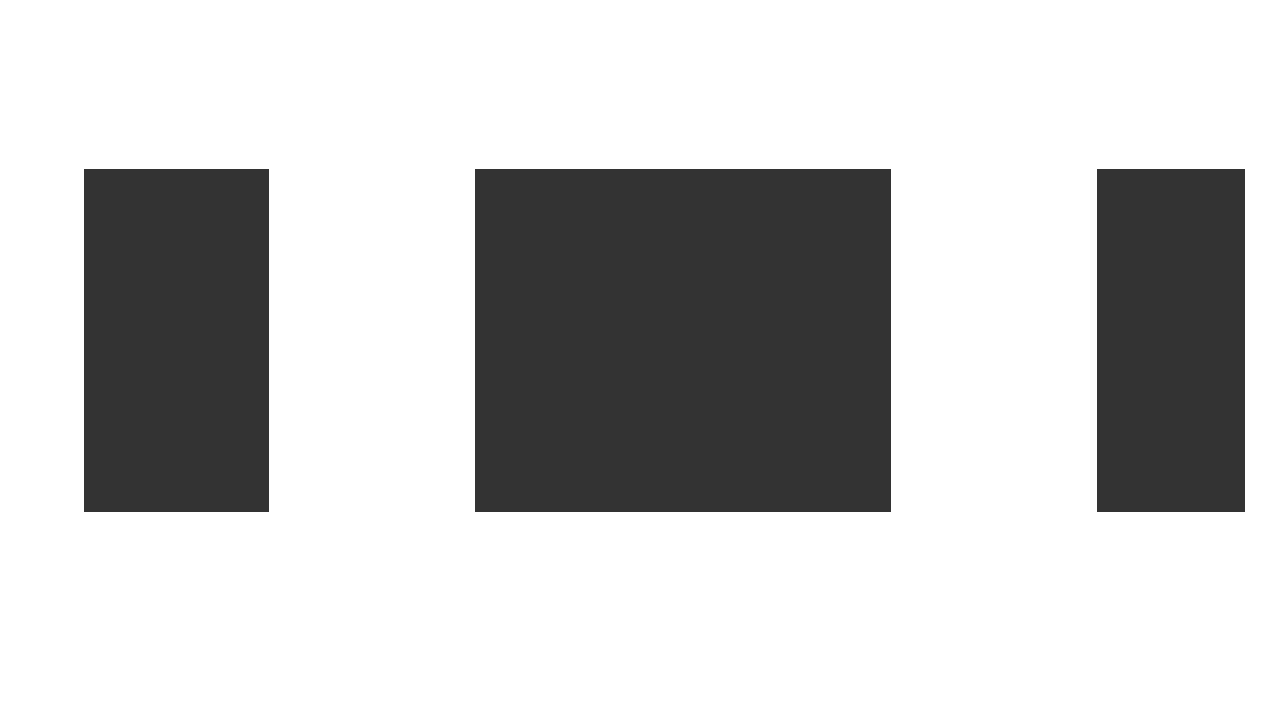}).}
\label{fig:results_individual_load}
\end{subfigure}
\caption{Load reductions and incentive rates for MARL-iDR (\protect\includegraphics[height=0.5em]{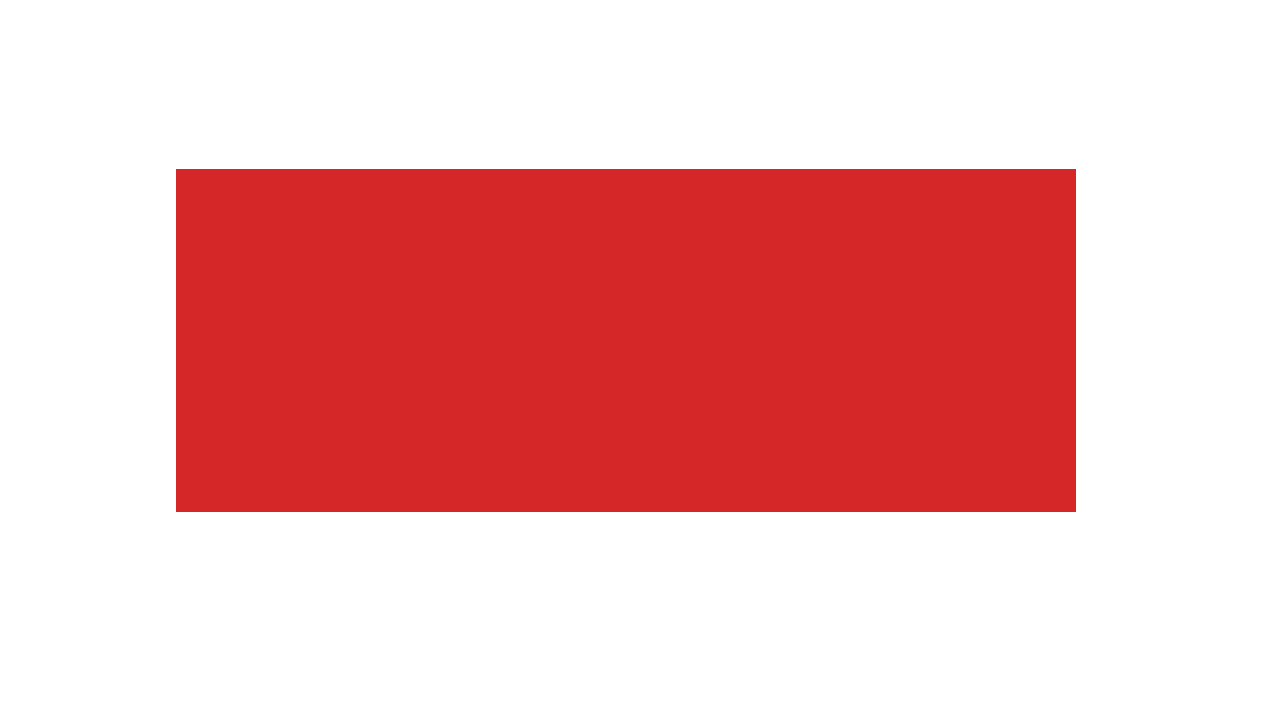}) compared to the myopic baseline (\protect\includegraphics[height=0.5em]{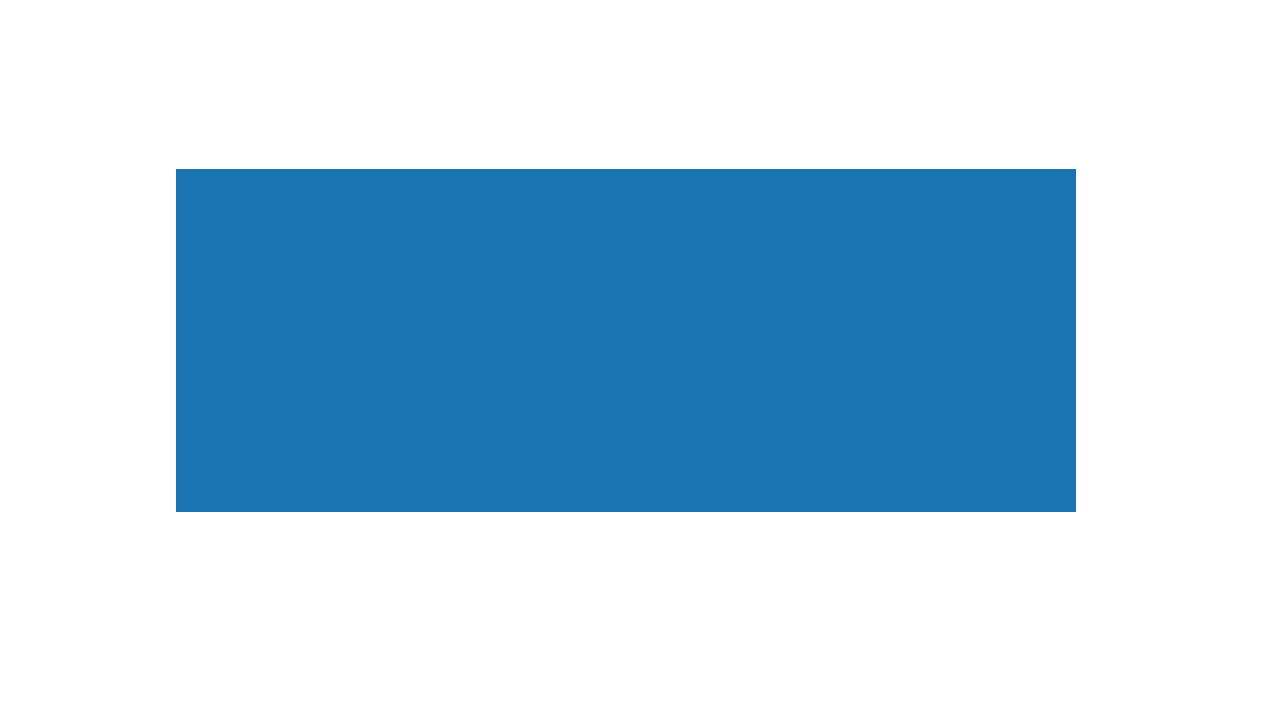}) and without DR (\protect\includegraphics[height=0.5em]{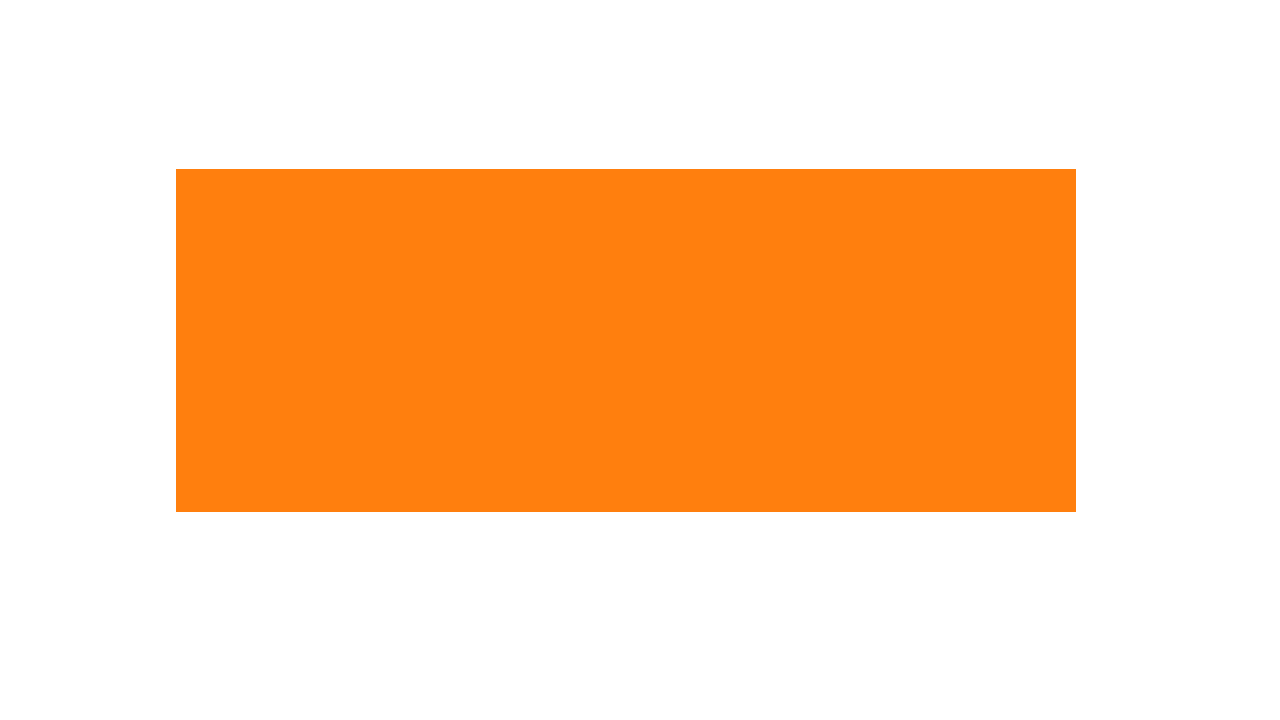}). 
}
\label{fig:results_load_reduction}
\end{figure}

MARL-iDR reduces loads during peak hours. The results are in Table \ref{tab:results}. The peak load and peak-to-average ratio (PAR) are significantly lower for MARL-iDR compared to the original load, i.e the case without DR. However, the myopic baseline reduces, even more, the peak load and PAR, slightly exceeding the target reduction with a total of $\SI{0.49}{\kWh}$, whereas this ``surplus consumption'' for MARL-iDR is $\SI{3.93}{\kWh}$.
MARL-iDR sometimes results in a second peak that exceeds the target reduction $k$. This behaviour of shifting the load to a second peak is known as the rebound effect \cite{Broka2020HandlingFramework}. Fig. \ref{fig:results_load_reduction} illustrates this behaviour, showing the impact of MARL-iDR on the load curve on July 1st. The aggregated load (Fig. \ref{fig:results_aggregated_load}) is unchanged before 14:30. During hours where the original load exceeds the target reduction, both MARL-iDR and the myopic baseline maintained the total load mostly below the target, by offering varying incentive rates. However, around 19:00, a second peak arises when using MARL-iDR. This second peak of $\SI{90.7}{\kW}$ is lower than the first, original peak $\SI{102.7}{\kW}$ but higher than the target. MARL-iDR does not offer incentives after the original peak. Hence the loads increase above the target (rebound effect). The myopic baseline does not suffer from the rebound effect and reduces below the target.
A similar pattern can be observed in individual households, see Fig. \ref{fig:results_individual_load} for the load curve of a selected household. Similar to the aggregated case, consumption is reduced significantly during peak hours but spikes right after 19:00.


\subsection{Dissatisfaction costs and appliance scheduling}
\begin{figure}
\begin{subfigure}{0.4\textwidth}
\begin{tikzpicture}
\begin{axis}[
        axis on top,
        width=\textwidth,
        height = 6em,
        scale only axis,
        enlargelimits=false, 
        ytick={0.5,1.5,2.5,3.5,4.5},     
        yticklabels={AC, EV, WM, DW, Dryer},
        xtick={12,14,16,18,20,22,24}, 
        xlabel={Time(h)},
        xmin=12,
        xmax=24,
        ymin=0,
        ymax=5,
        y label style={at={(axis description cs:0.1,.5)},anchor=south}, 
        ]        
	\addplot graphics[xmin=12,ymin=0,xmax=23.75,ymax=5] {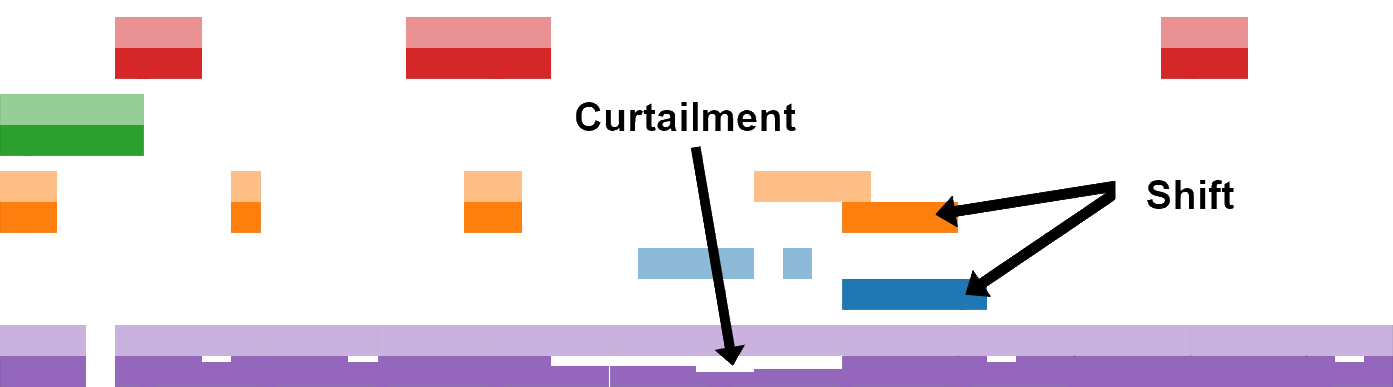};
  \end{axis}
\end{tikzpicture}
   \caption{Schedule with requests (light blocks) and assignments (dark blocks), block height is the curtailed AC level $q_j$.}
\label{fig:results_schedule}
\end{subfigure} 
\begin{subfigure}{0.4\textwidth} 
\begin{tikzpicture}
\begin{axis}[
        axis on top,
        width=\textwidth,
        height = 6em,
        scale only axis,
        enlargelimits=false, 
        ytick={-5, -2,5, 0, 2.5, 5},     
        xtick={12,14,16,18,20,22,24}, 
        ylabel={Reward},
        xlabel={Time(h)},
        xmin=12,
        xmax=24,
        ymin=-6,
        ymax=7,
        y label style={at={(axis description cs:0.1,.5)},anchor=south}, 
        ]        
	\addplot graphics[xmin=12,ymin=-6,xmax=23.75,ymax=7] {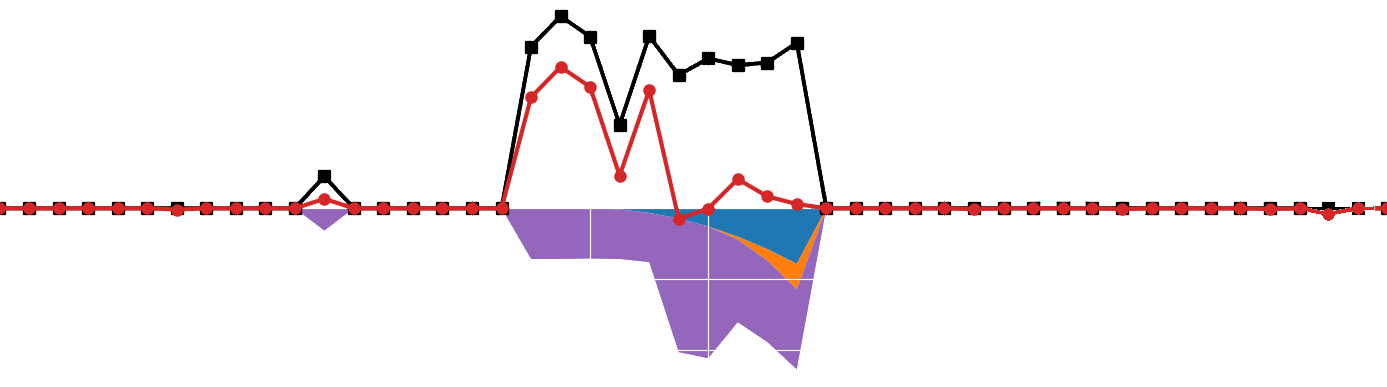};
  \end{axis}
\end{tikzpicture}
   \caption{Trading-off the two reward components of the PA, the financial reward $u$ minus the accumulated dissatisfaction $c$.}
\label{fig:results_dissatisfaction}
\end{subfigure}
\caption{Appliance scheduling of one household in (a) and in (b) is the corresponding financial reward $u$ (\protect\includegraphics[height=0.5em]{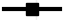}), total reward $r^{PA}$ (\protect\includegraphics[height=0.5em]{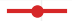}) and all components to the dissatisfaction $c_{EV}$ (\protect\includegraphics[height=0.5em]{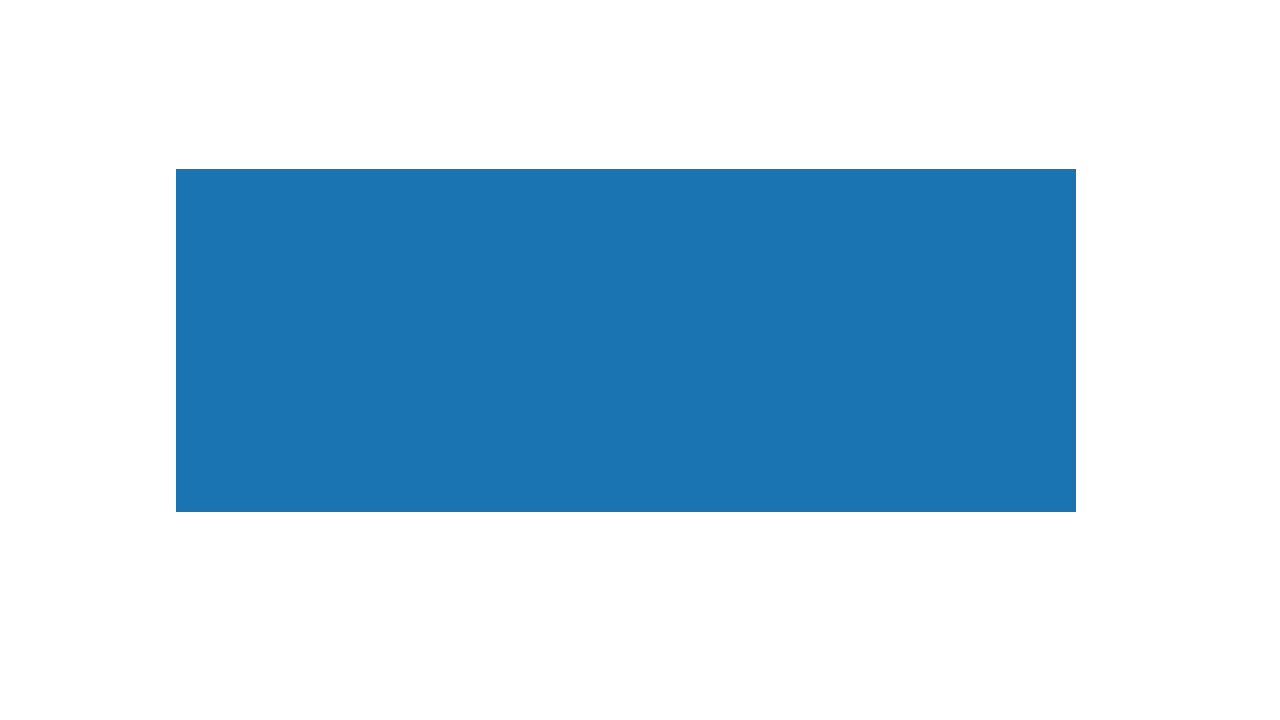}), $c_{WM}$ (\protect\includegraphics[height=0.5em]{Figures/images/withoutDR.png}), $c_{DW}$ (\protect\includegraphics[height=0.5em]{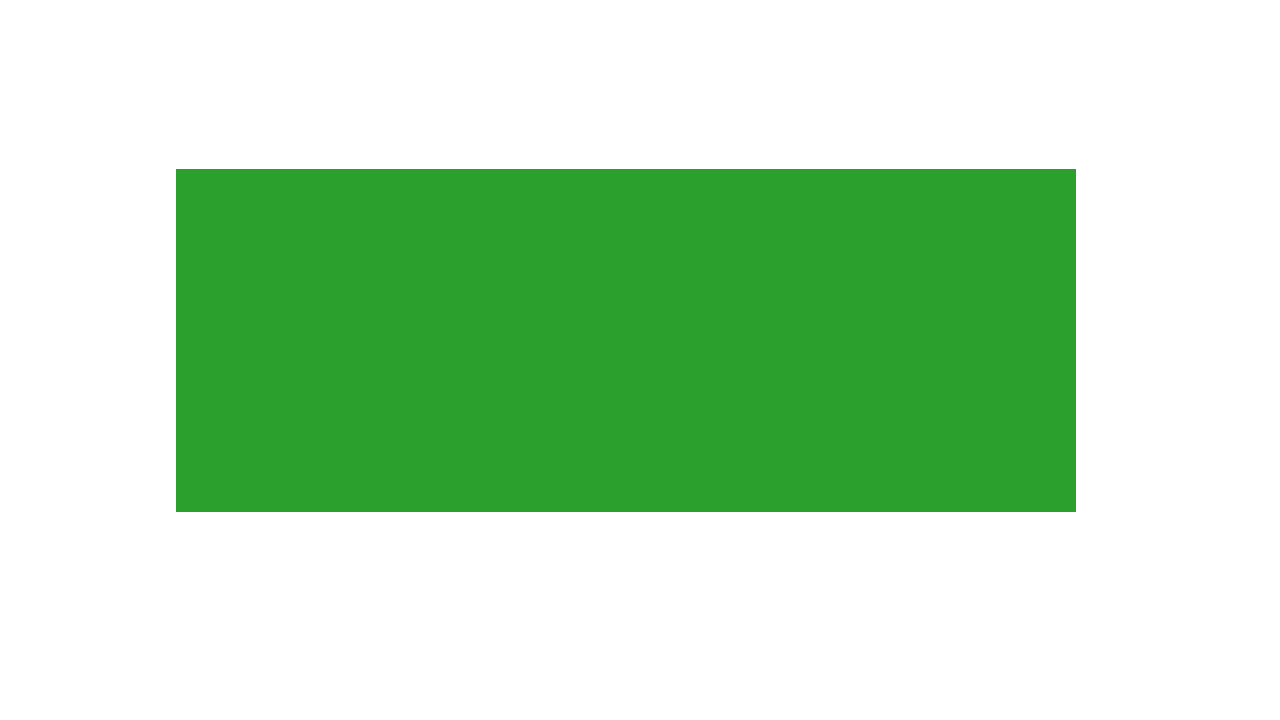}), $c_{Dryer}$ (\protect\includegraphics[height=0.5em]{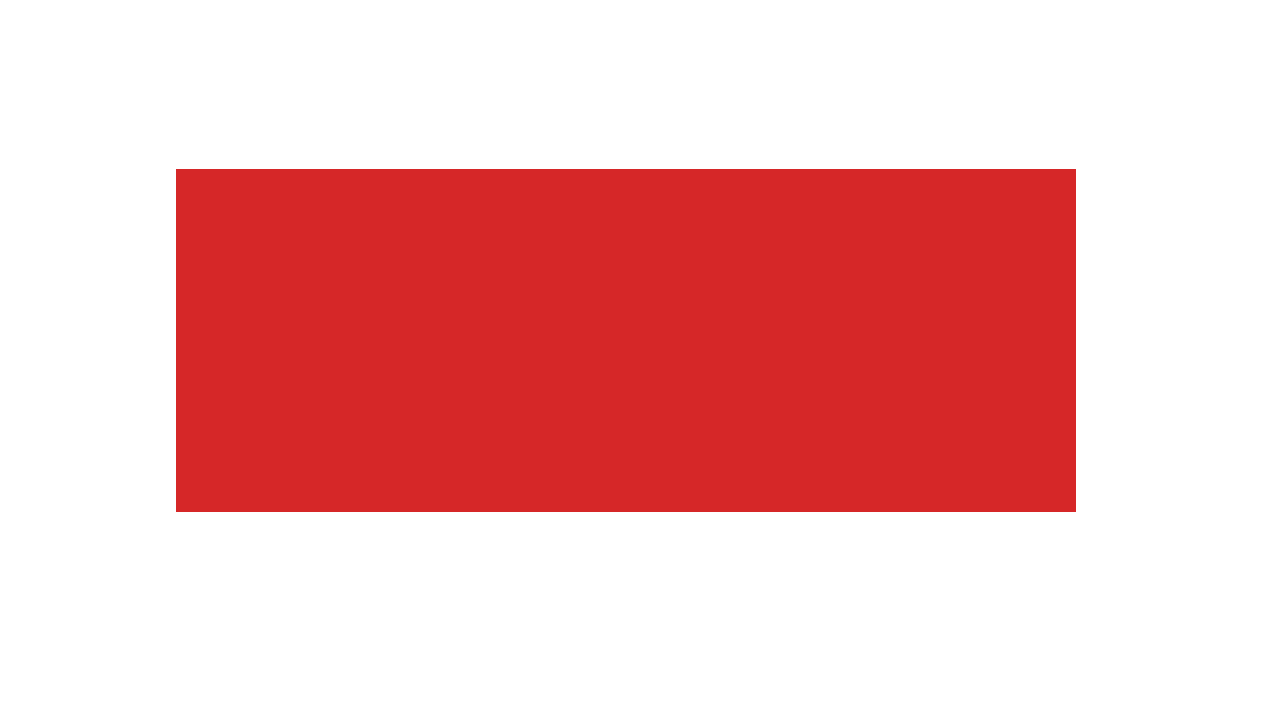}), $c_{AC}$ (\protect\includegraphics[height=0.5em]{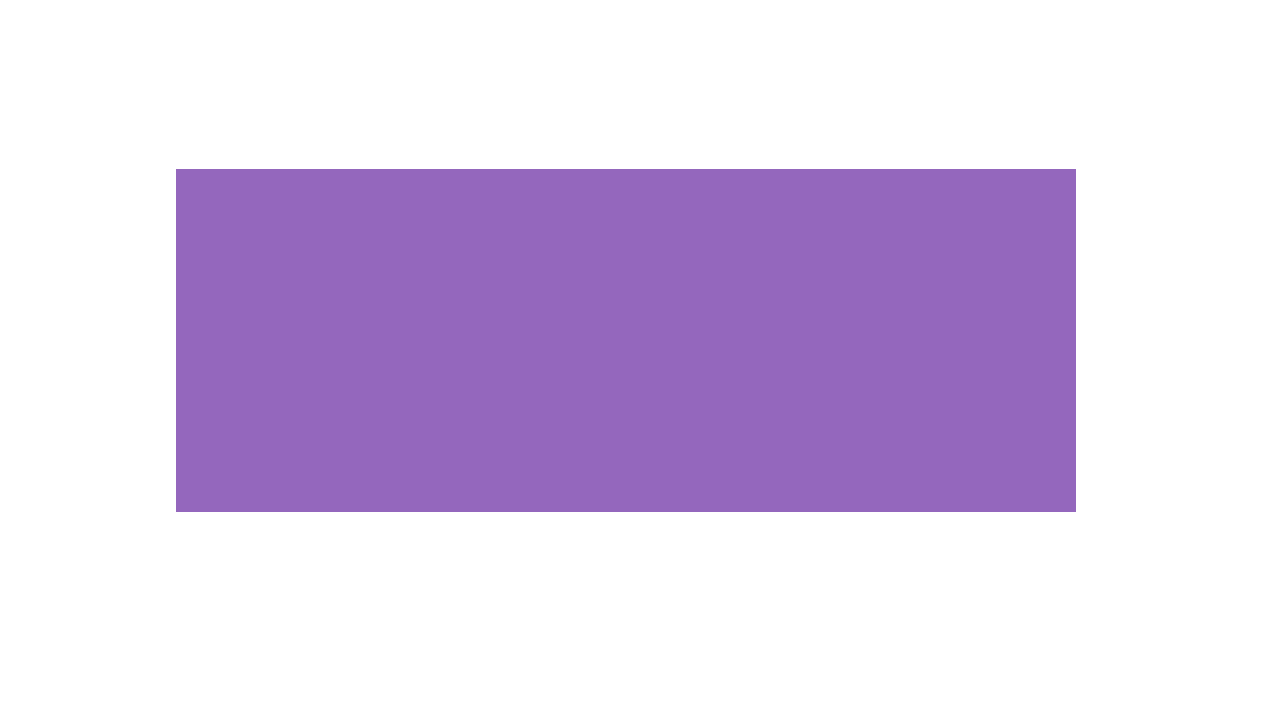}).\vspace{-1em}
}
\label{fig:results_single_household}
\end{figure}

The appliance schedule of the household from Fig. \ref{fig:results_individual_load} is in Fig. \ref{fig:results_single_household}. Fig. \ref{fig:results_schedule} shows each appliance's originally requested time step and the scheduled time step.
The washing machine and the EV delay for as long as incentives are offered. As soon as the incentive rate drops to $0$, shortly after 19:00, the agent schedules the washing machine and the EV.
The incentive also influences the AC. Between 16:30 and 19:00, the AC consumption is nearly halved. Fig. \ref{fig:results_dissatisfaction} shows the trade-off between incentives gained and dissatisfaction caused by rescheduling appliances. The dissatisfaction from postponing EV charging and the washing machine increases until 19.15.

\subsection{Preserving privacy}
The proposed MARL-iDR preserves privacy which is legally required. MARL-iDR outperforms any centralized scheduler at the AA (e.g., myopic baseline) as they require knowing all resident details, such as exact information about the reward function, to calculate the best responses. In practice, the aggregator must know participants' preferences and the state of their appliances to predict their response. With MARL-iDR the aggregator receives no information regarding the participant; only a single value-information is exchanged, the incentive rate. In addition, no information is exchanged between participants, ensuring participant privacy. 

\subsection{Economic analysis}


MARL-iDR trains local agents to balance the economics between the aggregator and all participants. The key metric in this balance is the incentive rate and the AA-designed parameters of the DR scheme.  

The incentive rates selected by the AA are shown in Fig. \ref{fig:results_incentive}. The shape of this rate curve matches the original aggregated demand curve in Fig. \ref{fig:results_aggregated_load}. MARL-iDR stops offering incentives after 19.00, and the myopic baseline stops after 20.00. The myopic baseline offers less or equal incentives optimizing the target reduction between 14.45 and 18.15, which results in lower financial rewards for the PAs.

\begin{figure}
\begin{subfigure}[t]{0.17\textwidth}
\begin{tikzpicture}
\begin{axis}[
        axis on top,
        width=\textwidth,
        scale only axis,
        enlargelimits=false, 
        ytick={0,2000,4000,6000},     
        xtick={0.1,0.3,0.5,0.7,0.9}, 
        ylabel={Incentive costs (\textcent)},
        xlabel={$\rho$},
        xmin=0.1,
        xmax=0.9,
        ymin=0,
        ymax=6000,
        y label style={at={(axis description cs:0.1,0.5)},anchor=south}, 
        ]        
	\addplot graphics[xmin=0.1,ymin=0,xmax=0.9,ymax=6000] {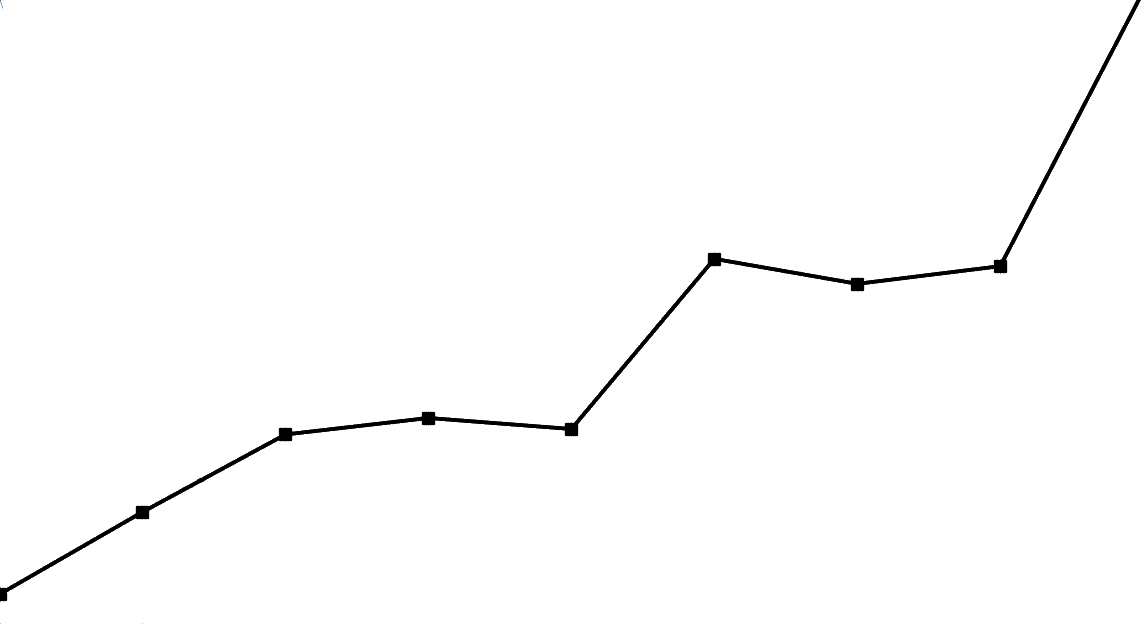};
  \end{axis}
\end{tikzpicture}
\caption{}
\label{fig:incentive}
\end{subfigure} \hspace{0.07\textwidth}
\begin{subfigure}[t]{0.17\textwidth}
\begin{tikzpicture}
\begin{axis}[
        axis on top,
        width=\textwidth,
        scale only axis,
        enlargelimits=false, 
        ytick={0,5,10,15,20},     
        xtick={0.1,0.3,0.5,0.7,0.9}, 
        ylabel={Surplus (kWh)},
        xlabel={$\rho$},
        xmin=0.1,
        xmax=0.9,
        ymin=0,
        ymax=20,
        y label style={at={(axis description cs:0.25,.5)},anchor=south}, 
        ]        
	\addplot graphics[xmin=0.1,ymin=0,xmax=0.9,ymax=20] {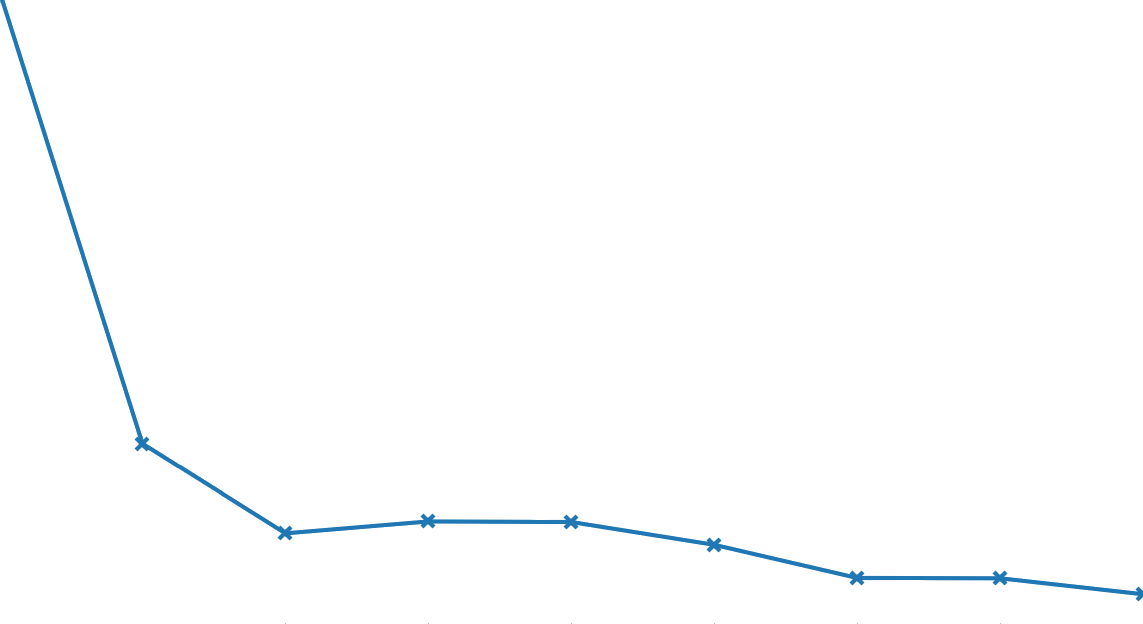};
  \end{axis}
\end{tikzpicture}
\caption{}
\label{fig:surplus}
\end{subfigure}
\caption{The aggregator balances financial cost (a) and surplus consumption (b) with the parameter $\rho$.}
\label{fig:results_rho_aa}
\end{figure}

The design of the program is important for its success toward a fair balance in financial costs and gains for all parties. The design (and the balance) is controlled at the AA by selecting the weighting factor $\rho$ in Eq. \eqref{eq:rAA}. The selection of $\rho$ ultimately determines households' willingness to participate in the DR program. A study on this parameter $\rho$ is in Fig. \ref{fig:results_rho_aa}. The larger $\rho$, the larger the punishment on surplus consumption and the smaller the incentive cost. For small values $\rho$, a significant consumption exceeds the target while only a little incentive is paid to participants. On the other hand, when $\rho$ is large, the aggregator tries to push the surplus consumption down to 0 by offering increasing incentives.

\subsection{Computational efficiency}
This study analyzes the computational times for MARL-iDR training and real-time deployment. The computation time during real-time deployment is an important criterion for the future needs of residential DR programs. The computation times of scheduling the appliances of a single household in a one-time step are compared with the baseline. A myopic baseline is a centralized approach to computing all possibilities before the PA can decide the optimal scheduling. The used implementation took, on average, $\SI{1.86}{\second}$, and for a large number of households, centralized optimization-based approaches are highly unsuitable, as research shows. However, as MARL-iDR is decentralized and the PAs can schedule appliances independently, the actual schedule can be computed in \SI{2}{\milli\second}. As the scheduling of appliances should be done in near real-time, MARL-iDR is very suitable as a real-time decision-making algorithm for real-time DR programs. The key advantage is that almost unlimited many PAs can be considered simultaneously. The time of training MARL-iDR with one AA and 25 PAs for $5000$ episodes is $\sim \SI{12}{\hour}$, which only has to be done once before deployment. 


The proposed decentralized MARL-iDR approach is very promising for future real-time DR programs as it scales to very large numbers of residents, making DR decisions in milliseconds while preserving privacy and balancing financial gains among participants in a fair way. However, MARL-iDR has limitations. As the AA makes decisions based on the current state of the environment and can not know if the current time step is before or after the peak as of the nature of MDPs, incentives were not placed to reduce the second peak. Hence, the myopic baseline outperformed MARL-iDR in reducing load. One way of solving this limitation could be to include the accumulated load reduction in the observation which requires further investigation. 
Another limitation is that the scheduler only considers requests for appliances at time step $t$, hence non-interruptible appliances may impede load reduction in future time steps when incentives may be higher. Operation times of time-shiftable appliances must be considered in the future to improve the potency of appliance scheduling. Finally, MARL-iDR is trained and validated in a period with relative high outside temperatures and large AC consumption (April to October). In the future, different characteristics should be considered to analyze the generalizability of the proposed method.

\section{Conclusion} \label{sec:conclusion}
In conclusion, this paper proposed a decentralized Multi-Agent Reinforcement Learning (MARL) approach to an incentive-based Demand Response (DR) program that addresses the key challenge of coordinating heterogeneous preferences and requirements from multiple participants while preserving their privacy and minimizing financial costs for the aggregator. The proposed approach was validated through case studies with electricity data from $25$ households. It was shown to effectively reduce the Peak-to-Average ratio (PAR) of energy consumption by $14.48$\% compared to the original PAR while fully preserving participant privacy. However, the MARL-IDR algorithm showed some rebound effects and did not always achieve the target reduction and the myopic baseline. The results of this case study demonstrate the proposed approach's potential to improve the electricity grid's efficiency and reliability. The novel Disjunctively Constrained Knapsack Problem optimization used to curtail or shift the requested household appliances based on the selected demand reduction makes this approach valuable to managing renewable energy resources and the growing electricity demand. Future work should address the rebound effect and improve the algorithm's performance.


\bibliographystyle{IEEEtran}
\bibliography{main.bib}

\end{document}